\documentclass[aps,prb,twocolumn,superscriptaddress,showpacs,floatfix]{revtex4-1}
\usepackage{graphicx}
  \graphicspath{{./figs/}}

\begin{document}

\title{Dynamics of Vortex Nucleation in Nanomagnets with Broken Symmetry}
\date{\today}
\author{Jaroslav T\'{o}bik}
\email{jaroslav.tobik@savba.sk}
\affiliation{Institute of
Electrical Engineering, Slovak Academy of Sciences,
D\'{u}bravsk\'{a} cesta 9, SK-841 04 Bratislava, Slovakia}

\author{Vladim\'{\i}r Cambel}
\affiliation{Institute of Electrical Engineering, Slovak Academy of Sciences, D\'{u}bravsk\'{a} cesta 9, SK-841 04 Bratislava, Slovakia}
\author{Goran Karapetrov}
\affiliation{Institute of Electrical Engineering, Slovak Academy of Sciences, D\'{u}bravsk\'{a} cesta 9, SK-841 04 Bratislava, Slovakia}
\affiliation{Department of Physics, Drexel University, 3141 Chestnut Street, Philadelphia, Pennsylvania 19104, USA}

\begin{abstract}
We investigate fundamental processes that govern dynamics of
vortex nucleation in sub-100 nm mesoscopic magnets. We focus on a
structure with broken symmetry - Pacman--like nanomagnet shape -
in which we study micromagnetic behavior both by means of a simple
model and numerically. We show that it is possible to establish
desired vortex chirality and the polarity of vortex core by
applying only quasi--static in-plane magnetic field along specific
directions. We identify the modes of vortex nucleation that are
very robust against external magnetic field noise. These vortex
nucleation modes are common among wide range of sub-100 nm magnets
with broken rotational symmetry.
\end{abstract}
\maketitle

Confinement leads to fundamental changes in physical behavior of
materials due to increased role of the surface. In mesoscopic
magnetic materials such changes in the energy landscape could lead
to novel magnetic spin configurations such as vortices.
Equilibrium properties of these new topological states are
governed by both the properties of the magnetic material and the
geometry of the object. On the other hand, confinement also leads
to distinct dynamic behavior of the new topological states, since
the available energy levels are very much limited. The transition
probabilities  between different states can thus be controlled by
careful engineering of the geometry of the mesoscopic object. Here
we show that by tailoring the geometry of the mesoscopic magnet
one can produce deterministic dynamic switching between well
defined degenerate topological states using only in-plane magnetic
pulses. We present a simple model that explains the mechanism of
the vortex nucleation and origin of robustness of the vortex
polarization. Confirmation of the model is accomplished by using
numerical simulations.

Controlled manipulation of magnetic domains in ferromagnet
nanostructures have recently opened opportunities for novel fast,
high-density, and low-power memories with novel architectures
\cite{Parkin,JVAC99_Ross,NatPhys_Novosad}. Any perspective
magnetic memory architecture, such as sub-micron non-volatile
magnetic memory, has to have (1) a well-defined switching field
used to set the memory bits, and (2) a reproducible switching
behavior using simple sequence of external magnetic field pulses.
Therefore, the dynamics of the switching between different ground
states has to be understood in detail.

Recent advances in fabrication technology at nano-scale have
enabled studies of magnetic systems that are well defined in all
three dimensions on a nanometer length scale ($<$ 100
nm)\cite{Chung_2010,JPD_Lau}. The size reduction of nanomagnets
leads to novel spin topological states such as vortex state, C-
state, S-state, flower state, etc. \cite{JPD_Cowburn}, and to
simplified transitions between these states in external magnetic
field. The transition between ground states in such nanomagnets is
of fundamental importance \cite{Brown}. It is governed by
competition between the magneto-static energy and exchange energy,
and it is influenced by the magnetic material used and by the
choice of the nanomagnet shape. For example, the magnetization of
disks in zero field can be oriented in-plane, out-of-plane, or
vortex state can be created, depending on the disk diameter and
thickness \cite{Chung_2010,PRL_Albuquerque}. In the disk the
flux-closure magnetic state reduces the long range stray fields,
i.e. reduces the magnetostatic interaction between neighboring
disks. Therefore, such disk magnetic systems have a potential for
high-density magnetic storage elements, with bits represented by
chirality and polarity of a basic vortex state.

In sub-micron sized disks four possible states have to be
controlled. It was shown experimentally
 that in the nanomagnets
with broken rotational symmetry, chirality can be controlled
easily by in-plane field of selected direction
\cite{APL79_Schneider, JAP97_Taniuchi,
JAP99_Vavassori,PRB81_Jaafar}. At the same time the polarity of
the vortex core, which represents the second bit, can be
controlled by an out-of-plane magnetic field \cite{PRB81_Jaafar},
spin--polarized current \cite{Nature_Yamada} , high-frequency
in-plane magnetic field \cite{PRB76_Lee}, or by an in-plane
magnetic pulse of precisely defined amplitude and duration
\cite{PRB80_Antos}.

In our previous work we have proposed a new prospective shape of a
nanomagnet with broken symmetry which permits control of chirality
and polarity bits by the application of the in-plane field only
\cite{PRB84_Cambel}.
In this paper we analyze the mechanisms that establish specific
chirality and polarity values in such Pacman-like (PL) nanomagnet
by taking a closer look at energies and dynamics that govern the
switching processes. We show that the polarity and the chirality
of the vortex core nucleated in the decreasing in-plane magnetic
field is implicitly defined by the direction of the magnetic field
with respect to missing sector of the PL nanomagnet.

\begin{figure}
\includegraphics[width=0.70\columnwidth]{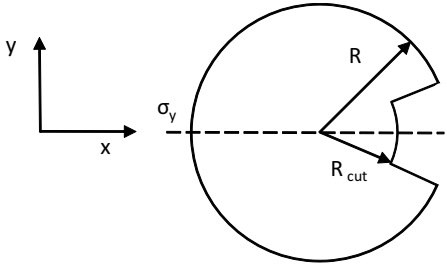}
\caption{\label{fig-PLstructure} Geometry of Pacman--like
nanomagnet. The structure is symmetric with respect to reflection
plane $\sigma_y$. }
\end{figure}

We consider a magnetic dot of cylindrical shape. To construct a PL
structure it is necessary to remove an outer sector that is
$45^\circ$ wide, and has $1/3$ of the disk radius (see Fig.
\ref{fig-PLstructure}). Due to symmetry analysis that will be
presented later, we choose the orientation of $x,z$ axes such,
that they define mirror symmetry plane $\sigma _y$, which leaves
PL object invariant. Another symmetry operation is mirroring
through the plane $x,y$ noted in following text by $\sigma_z$.

First, we define polarity $\vec{\pi}[\vec{f}]$ and
chirality $\vec{\chi}[\vec{f}]$ vectors as functionals of an arbitrary vector
field $\vec{f}$:
\begin{eqnarray}
\vec{\pi}[\vec{f}]&=&\frac{1}{\Omega}\int \vec{f}(\vec{r}) d\Omega \label{eq_pch_def}\\
\vec{\chi}[\vec{f}]&=&\int \vec{r}\times
\left( \vec{f}(\vec{r})-\vec{\pi} \right) d\Omega \nonumber
\end{eqnarray}
Polarity is just simple average value of the field, while
chirality resembles the definition of momentum of quantity
$\vec{f}$ in classical mechanics. Subtraction of polarity in the
expression for chirality is necessary due to chirality invariance
with respect to the origin coordinate system choice.
Mostly we are interested in $z$ component of polarity and chirality. 
Integration domain $\Omega$ is over the volume of the PL
nanomagnet.

\begin{figure}
\includegraphics[angle=-90, width=0.99\columnwidth]{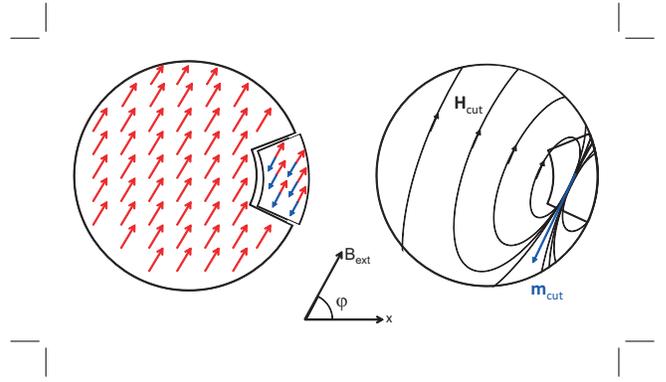}
\caption{ \label{fig_chirality_explanation} (a) The magnetization
of the Pacman--like nanomagnet is a superposition of the uniform
magnetization of a full disk (red arrows) and the magnetization of
the missing sector that is equal and opposite to the one of the
full disk. (b) The sum of the compensation moments in the sector
creates a dipole $m_{cut}$ which asymmetrically interacts with the
local magnetization in the nanomagnet.}
\end{figure}

Let's consider that the nanomagnet is placed in a strong in--plane
 magnetic field that has an angle $\varphi$ with
the $x$-axis. To emulate magnetic response by the missing sector,
we can consider the PL nanomagnet as a superposition of a full
disk and a set of microscopic magnetic moments in a removed
sector. These additional moments have to have the same value and
to be oriented in opposite direction to the magnetic moments in
the disk (Fig. \ref{fig_chirality_explanation}). In the first
approximation all microscopic moments are parallel. Neglecting
higher than dipolar moments, missing sector behaves as a dipole
with a moment $\vec{m}_{\text{cut}}$ positioned in the center of
mass of the sector $\vec{r}_{\text{T}}$:
\begin{equation}
\vec{m}_{\text{cut}}=-\int \vec{M} d\Omega'\quad
\vec{r}_{\text{T}}=\frac{1}{\Omega '}\int \vec{r} d\Omega'.
\label{eq_dipole_removed}
\end{equation}
Minus sign reflects that dipoles of opposite orientation have to
be added in order to eliminate dipoles in the missing sector (see
Fig. \ref{fig_chirality_explanation}). The integration is over the
volume of the missing sector $\Omega'$.

To calculate the magnetic field of PL nanomagnet consider first
the full disk. In magnetic fields exceeding the saturation field
the magnetic polarization $\vec{M}$ is parallel to the direction
of the applied field $\vec{H}_{\text{ext}}$ throughout the disk
volume. Internal magnetic field $\vec{H}$ is also uniformly
oriented in parallel with $\vec{M}$. As small piece of material is
removed, all the fields change slightly. To correct the internal
magnetic field, the field of magnetic moment
$\vec{m}_{\text{cut}}$ given by equation (\ref{eq_dipole_removed})
has to be added to the originally homogeneous internal field of
the full disk. The removed part thus creates a dipole which
induces magnetic field with non-zero chirality $\chi[\vec{H}]$, if
$\varphi\neq 0^\circ, 180^\circ$. Since dipoles partially follow
the field orientation, the non-zero chirality of $\vec{M}$ is
expected.

Next, we explain qualitatively the mechanism which determines
vortex core polarity. Taking into consideration magneto-statics
only, the state with positive polarity $\pi[\vec{M}]$ is
energetically equivalent to the state with negative polarity
$-\pi[\vec{M}]$, if no external field in $z$ direction is applied.
That can be easily seen by writing the total energy functional:
\begin{eqnarray}
E&=& \frac{\mu_0}{4\pi} \int \int  \left(
\frac{\vec{M}\cdot\vec{M}'}{|\vec{r}-\vec{r}'|^3} -\frac{3\vec{M}
\cdot(\vec{r}-\vec{r}')
\vec{M}'\cdot(\vec{r}-\vec{r}')}{|\vec{r}-\vec{r}'|^5}
\right) d\Omega d\Omega' \nonumber \\
&+&A\int (\nabla \vec{M})^2 d\Omega-\int
\vec{H}_{\text{ext}}\cdot\vec{M} d\Omega \label{eq-functional}
\end{eqnarray}
Changing the sign of $M_z$ does not change first two integrals of
the equation (\ref{eq-functional}), since these two parts of the
energy functional are quadratic in $M_z$ and its derivative,
respectively. The only linear part in $M_z$ is in the third
integral. But a change of sign of $M_z$ would not influence the
total energy, because $H_{\text{ext}}$ has no $z$ component.
Integration domain $\Omega$ is not changed by reversing $z$,
because symmetry operation $\sigma_z$ - reflection from the plane
$xy$ transforms PL nanomagnet to itself.

The final state of vortex core polarization is determined by
magnetization dynamics. Time evolution of magnetization is
described by phenomenological Landau-Lifshitz-Gilbert (LLG)
equation:
\begin{equation}
\frac{\partial \vec{M}}{\partial t}=-\gamma \vec{M}\times\vec{H}_{\text{eff}}+
\alpha\left( \vec{M}\times \frac{\partial \vec{M}}{\partial t}
\right)
\label{eq_LLG}
\end{equation}

\begin{figure}
\includegraphics[width=1.02\columnwidth]{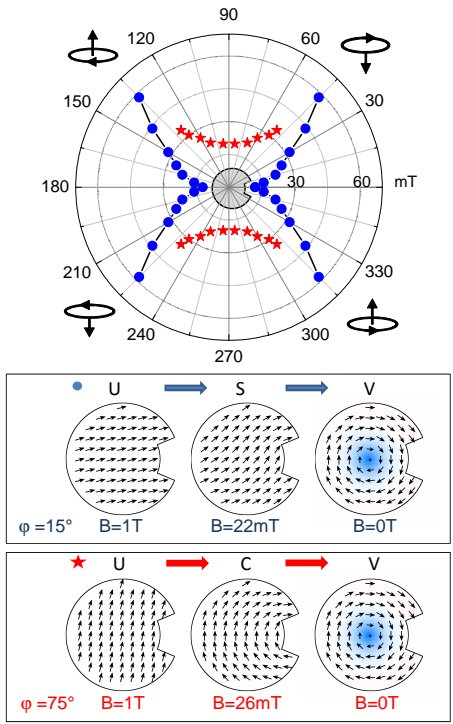}
\caption{ \label{fig_bnuc_angular} Angular dependence of the
vortex nucleation process. Top: angular dependence of vortex
nucleation field. Direction of the nucleated vortex polarity and
chirality are also indicated for each quadrant. Bottom: two
different processes of vortex nucleation depending on initial
magnetization direction with respect to PL's symmetry plane
($\varphi=15\circ \text{ and } \varphi=75\circ$): from uniform
magnetization state the magnetization transitions to S-shape
(dots) or C-shape (stars) configurations and equilibrates to a
vortex state with specific polarity and chirality.}
\end{figure}

Here we show that this equation itself contains polarity symmetry
breaking mechanism. First, we apply strong external in-plane field
in direction that has an angle $\varphi$ with the $x$-axis
(Fig.\ref{fig_chirality_explanation}). Then we slowly
(adiabatically) decrease the external field amplitude to the level
that is just above the vortex nucleation field. By adiabatic field
change we mean that the change of the external field with time is
so slow, that the energy dissipation keeps the system very close
to the local minimum at all times. Then $\frac{\partial
\vec{M}}{\partial t}\approx0$ everywhere. Any dynamics means also
dissipation of energy due to term that is proportional to $\alpha$
in equation (\ref{eq_LLG}). Therefore, at local minima we also
have $\vec{H}_{\text{eff}}\approx 0$, otherwise it is not possible
to satisfy equation (\ref{eq_LLG}) with vanishing left side. Now,
having $\vec{H}_{\text{ext}}$ just above nucleation field, we
decrease the external field by a small value $\Delta \vec{H}$. To
first approximation the effective field is
$\vec{H}_{\text{eff}}=\Delta \vec{H}$. When looking at the
dynamics of local magnetic moments shortly after the external
field is decreased, we can neglect the damping term in the LLG
equation, since $\alpha \ll 1$. The dynamics of $M_z$ is then
governed by the equation
\begin{equation}
\frac{\partial M_z}{\partial t}=-\gamma\left(M_x\Delta H_y-M_y\Delta H_x \right).
\label{eq_polarity_born}
\end{equation}

\begin{figure}
\includegraphics[width=0.99\columnwidth]{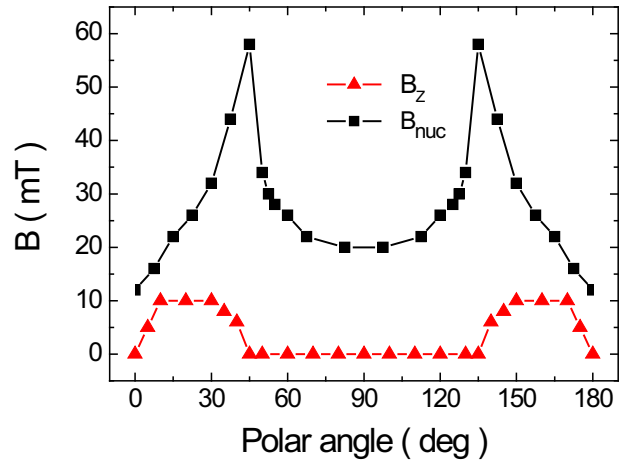}
\caption{ \label{fig_B_crit} Angular dependence of the in-plane
vortex nucleation field $B_{\text{nuc}}$ and threshold
out-of-plane field $B_{\text{z}}$ necessary to reverse the
polarity of the entering vortex.
}
\end{figure}

The right-hand side of the equation (\ref{eq_polarity_born}) is
not zero locally, nor it is in average, due to asymmetry of the PL
nanomagnet with respect to the direction of the applied field (if
$\varphi\neq0^\circ, 180^\circ$). If the actual value of the
external field is lower than the nucleation field, the non-zero
polarity of the nanomagnet will emerge. Paths towards two
equivalent minima characterized by $\pm M_z$ are energetically
equivalent, but change of external field with time is biasing the
time evolution of magnetization towards configuration given by eq.
(\ref{eq_polarity_born}).

Confirmation of the above model approximation can be obtained by
numerical simulations. We have performed numerical simulations of
PL nanomagnet using OOMMF software package\cite{OOMMF}. Parameters
used in the simulation are: outer radius $R=35$ nm, thickness (in
$z$-direction - not shown) $h=$ 40nm. Material used in the
calculations was Permalloy
$\text{Ni}_{\text{80}}\text{Fe}_{\text{20}}$ (Py), with following
material parameters: exchange constant
$\text{A}=13\times10^{-12}\text{J/m}$ and saturated magnetization
$M_s=8.6\times10^{5}\text{A/m}$.

In Figure \ref{fig_bnuc_angular} we show the dependence of the
nucleation-field amplitude on the applied field direction. The
nucleation field is defined as the applied magnetic field at which
the non-zero $z$ component of the magnetization polarity
$\pi(\vec{M})$ appears. External field is adiabatically decreasing
from 150 mT to zero in selected direction. By adiabatic change we
mean repeated process of decreasing field by 2 mT step, followed
by full system relaxation.

We would like to note the symmetry properties of PL nanomagnet. In
Fig.~\ref{fig_bnuc_angular}(top) each quadrant corresponds to a
vortex ground state of the nanomagnet with specific chirality and
polarity shown in corresponding corners. The nanomagnet relaxes
into that remanent state from a uniform magnetization along an
angle within the specific quadrant. As can be seen in
Fig.\ref{fig_bnuc_angular}, angular dependence of the nucleation
field can be reconstructed from the dependence for $\varphi \in
[0;\frac{\pi}{2}]$ by inversion and reflection through $xz$ plane
$\sigma_y$. Inversion symmetry of the graph shown in
Fig.\ref{fig_bnuc_angular} is the consequence of the time-reversal
symmetry. The reflection symmetry with respect to $xz$ plane shown
in Fig.\ref{fig_bnuc_angular} is related to reflection from
$\sigma_y$ - geometrical operation that transforms PL nanomagnet
to itself.

The simulation results show the existence of \emph{two distinct
vortex core nucleation regimes} (Fig.\ref{fig_bnuc_angular}
(bottom)). For large angles ($\varphi \in [50^\circ ;90^\circ]$) -
vortex nucleates from C-state magnetization pattern. This form of
nucleation is not robust in the sense that even small out-of-plane
field $B_z\simeq 1$mT is sufficient to alter the resulting
polarity of the nucleated vortex along the direction of applied
field $B_z$. Instead, for small angles ($\varphi \in [0^\circ ;
48^\circ ]$) - the vortex nucleation path is different. Just above
the vortex nucleation field, the magnetization of PL nanomagnet
forms an S-state. This configuration consists of two regions with
opposite signs of the curvature of field
lines~\cite{note_curvature}. Meanwhile, there is only one
curvature of field lines just below the vortex nucleation field.
The process of vortex core nucleation in this case involves {\em
reversal of magnetic moments}~ in a part of the nanomagnet. This
reversal process proceeds through an out-of-plane motion of the
local magnetic moments, resulting in robust vortex core
polarization despite the presence of small external fields in $z$
direction. In figure \ref{fig_B_crit} we show the angular
dependence of the maximum external field $B_z$ for which the PL
nanomagnet is able to sustain nucleation of vortex polarity
opposite to the direction of applied external field.

The C and S-shapes of magnetization can be explained by the
position of perturbing dipole $\vec{m}_{\text{cut}}$. As can be
seen from the Fig.\ref{fig_chirality_explanation}, PL nanomagnet
is divided into two domains with opposite sign of magnetic field
circulation generated by $\vec{m}_{\text{cut}}$. The size of this
two domains are determined by orientation of
$\vec{m}_{\text{cut}}$. However, exchange interaction tends to
align local moments in parallel, thus there exists a critical
angle (around $48^\circ$ in our geometry), beyond which the region
with minor curvature does not exist. Detailed energy balance
between exchange and cavity (sector) demagnetization determines
the scenario of vortex nucleation and its eventual robustness with
respect to external perturbation.

To summarize, in this work we provide simple arguments that
elucidate the origin of driving mechanisms for nucleation of
magnetic vortex with controlled chirality and polarity. We have
also found the regime of PL nanomagnet operation in which the
final vortex state is independent on weak disturbing external
field. This is a promising finding to consider if using PL
nanomagnet as a memory element in bit-patterned media or as a
generator of magnetic vortices of desired polarity and chirality
for microwave applications. Weak interaction among the elements as
well as robustness to small external field perturbations makes PL
nanomagnet very suitable for operation.

Finally we note, that the sub-100 nm PL nanomagnet is not the only
unique design offering control of chirality and polarity by
in-plane magnetic field. Qualitatively similar results are
obtained in simulations for different sizes and shapes of the
missing sector. According to our simple model, the necessary
ingredients are symmetry of the object, the demagnetization field
strength of removed part and shape anisotropy induced by removed
part. Robustness of vortex polarity against $B_z$ is based on
vortex -- anti-vortex annihilation during vortex core nucleation.

\begin{acknowledgments}
This work has been supported by the project CENTE II, Research \& Development
Operational Program funded by the ERDF, ITMS 26240120019, and by VEGA  2/0037/12.
\end{acknowledgments}


\begin{thebibliography}{99}
\bibitem{note_curvature}
To be precise, we mean curvature of field lines projected onto
$xy$ plane, since the concept of curvature with sign is meaningful
for curves in two dimensions.
\end{thebibliography}
\end{document}